# Modeling the Behavior of Reinforced Concrete Walls under Fire, Considering the Impact of the Span on Firewalls

\* Nadia Otmani-Benmehidi
Civil engineering, Laboratory: LMGE,
University Badji Mokhtar
Annaba, Algeria
benmehidi_nadia1@yahoo.fr

Meriem Arar
Civil engineering, Laboratory: LMGE,
University Badji Mokhtar
Annaba, Algeria
maria.2a21@hotmail.fr

Imene Chine
Civil engineering, Laboratory: LMGE,
University Badji Mokhtar
Annaba, Algeria
chineimene@yahoo.com

*Abstract* — Numerical modeling using computers is known to present several advantages compared to experimental testing. The high cost and the amount of time required to prepare and to perform a test were among the main problems on the table when the first tools for modeling structures in fire were developed. The discipline structures-in-fire modeling is still currently the subject of important research efforts around the word, those research efforts led to develop many software. In this paper, our task is oriented to the study of fire behavior and the impact of the span reinforced concrete walls with different sections belonging to a residential building braced by a system composed of porticoes and sails. Regarding the design and mechanical loading (compression forces and moments) exerted on the walls in question, we are based on the results of a study conducted at cold. We use on this subject the software Safir witch obeys to the Eurocode laws, to realize this study. It was found that loading, heating, and sizing play a capital role in the state of failed walls. Our results justify well the use of reinforced concrete walls, acting as a firewall. Their role is to limit the spread of fire from one structure to another structure nearby, since we get fire resistance reaching more than 10 hours depending on the loading considered.

## I. INTRODUCTION

A structure must be designed and calculated so that it remains fit for use as planned. It must resist to different degrees of reliability during execution as well as that during service. Finally, the structure must have adequate durability regarding the maintenance costs. To meet the requirements outlined above, we must: choose the materials appropriately, define a design and appropriate dimensioning. For this purpose, it is imperative to provide rules specific to each country. Various researches were performed by experts in the field of fire, to find out the behavior of the structures; as examples the separations and the bearer elements (concrete column, steel column…) of a building during a fire; which has developed fire rules. Regarding the fire behavior of bearing walls, among the authors working in this field, we mention Nadjai A [1], who performed a numerical study validated by an experimental investigation on masonry walls. Also, Cheer-Germ Go and Jun-Ren Tang [12] presented an experimental investigation.

Our work presents a contribution to the study of the behavior of reinforced concrete walls, cast in place, exposed to fire, belonging to a residential building. These walls were studied under the rules of wind and earthquake, by engineers in preparation for their final project in study. The building is composed of a ground floor + 9 floors, located in the Prefecture of Annaba (Algeria) [2].

In a fire situation the temperature building rises as a function of the material combustibility and the present oxygen. The fire causes degradation in characteristics of the material, a deformation in structural elements, and cracks will appear; finally, the structure is in ruin. In order to prevent those phenomena and to minimize the spread of the disaster with controlling it as quickly as possible, we can use the firewall in buildings.

In this paper we will study four concrete walls; two walls with a section 20×470 cm$^2$ reinforced with bars of Ø10 and two other walls having a section of 20×350 cm 2 (reinforced with Ø12). We consider a strip of 20 cm to reduce the work. The thermal loading is defined by the standard fire ISO 834[3]. Three walls are exposed to fire on one side; the fourth wall is exposed on two of its sides. The mechanical loading (i.e. compressive load and moment) exerted on the walls in question was taken from a study conducted at cold.

The thermal analysis gives the temperatures at every moment and at every point of the walls. These temperatures were used in the mechanical analysis. For the thermal analysis and the mechanical analysis we used the software Safir[4]. This software was developed by Franssen J M [4] in Belgium at the University of Liege, performed for the thermal and mechanical study of structures subjected to fire, taking into account the material and geometrical nonlinearity and large displacements. Rules Relating to concrete Firewalls

*A. Mechanical behavior relating to concrete: the Eurocode 2 model*

The division of the macroscopically measurable strains in heated concrete into individual strain components is done in the EC2 according to Eq (1)[7][14]:

$$\varepsilon_{tot} = \varepsilon_{th} + \varepsilon_{\sigma} + \varepsilon_{tr} + \varepsilon_{cr} \quad (1)$$

where $\varepsilon_{th}$ is the free thermal strain, $\varepsilon_{\sigma}$ is the instantaneous stress-related strain, $\varepsilon_{tr}$ is the transient creep strain and $\varepsilon_{cr}$ is the basic creep strain.

The mechanical strain is the sum of the instantaneous stress-related strain and the transient creep strain.

$$\varepsilon_{tot} = \varepsilon_{th} + \varepsilon_{m} + \varepsilon_{cr} \quad (2)$$

where $\varepsilon_m$ is the mechanical strain.

In implicit models, the stress is directly related to the mechanical strain, without calculation of the transient creep strain. In the EC2 model, the relationship at a given Temperature T between the stress and the mechanical strain is given for the ascending branch by Eq (3).

$$\frac{\sigma}{f_{c,T}} = \frac{3\varepsilon_m}{\varepsilon_{c1}\left(2 + \left(\frac{\varepsilon_m}{\varepsilon_{c1}}\right)^3\right)} \quad (3)$$

For more details we invite the lector to see [7].

*B. Firewalls with elements in cellular concrete*

We take as an example, a firewall [5] composed of concrete columns of 45×45 cm and panels of 600×60×15 cm (Posed in front or between the columns) presents a degree firewall equal to 4 hours. We must also note that the PV CSTB n° 87-25851 dated 11 /07 /95 precise that : "an experimental wall of element with reinforced cellular concrete of 15 cm thickness with a nominal density of 450 KG / m$^3$ mounted on flexible joints, has a degree firewall of 4 hours". Depending on the thickness, the limit height of wall is:
Wall thickness 15 cm corresponds to height: H = 17 m
Wall thickness 20 cm corresponds to height: H = 22 m
Wall thickness 25 cm corresponds to height: H = 28 m

As a first approximation, the degree of a firewall composed of solid panels with pre-cast concrete can be deduced from simplified rules, coming from the norm P 92-701 [6] expressed in Table I. These rules concern the walls with mechanical slenderness at most equal to 50 and are valid for a wall exposed to fire on one or two sides.

The concrete cast in place can be used to make firewalls. Implementation of these structures must respond to rules and code of calculation which concerned them ( DTU fire concrete) [6].

TABLE I. DEGREE FIREWALLS

|  | Degree CF | 1/2h | 1h | 1h30 | 2h | 3h | 4h |
|---|---|---|---|---|---|---|---|
| Bearing wall | Depth (cm) | 10 | 11 | 12 | 15 | 20 | 25 |
| Separating wall | Depth (cm) | 6 | 7 | 9 | 11 | 15 | 17.5 |

*C. Fire walls according to Eurocode2*

In section 5.4.3 of Eurocode 2[6], it is recommended that the minimum thickness for normal weight concrete, should not be less than:
200mm for unreinforced wall
140 mm for reinforced load-bearing wall
120 mm for reinforced non load bearing wall

*D. Fire walls according to our numerical study*

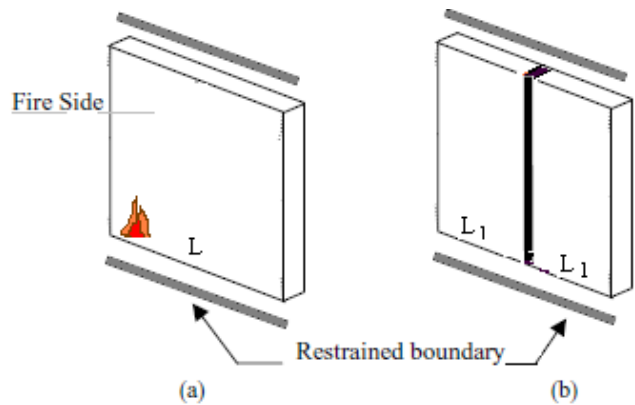

(a) Length of wall superior to 3.5 m
(b) The span wall must be reduced with forecast column

According to results of numerical study; we recommend to add column, when length of the fire wall exceed 3.5 m, to reduce the span wall.

$$1,5 \leq L_1 \leq 3,5$$

$$K_r = L_1/L ,$$

$$\rightarrow K_r = 3,5/L \quad (4)$$

$$L_1 = K_r \cdot L \quad (5)$$

$K_r$: factor of reduction
$L_1$: reduced length of wall [m]
L: initial length of wall [m]
This recommendation should be added in the Eurocode.

## II. MODELING OF WALLS

To begin the numerical study, it is necessary to model the walls considered. The Table II defines the geometrical characteristics and the loads. The reinforcement of each type of wall was calculated according to [2]. Walls: "Mu 20" and "MuF 2O" don't have the same thermal load. The first is subjected to normalized fire ISO834 [3] on one side, for the second one, we apply the fire on two sides. They contain reinforcements Ø 10 spaced with 20 cm "Fig. 1". They have the same section of $20 \times 470 cm^2$. Concerning the mechanical loading each wall is submitted to its ultimate moment (M-ult) and its ultimate compressive load (N-ult).

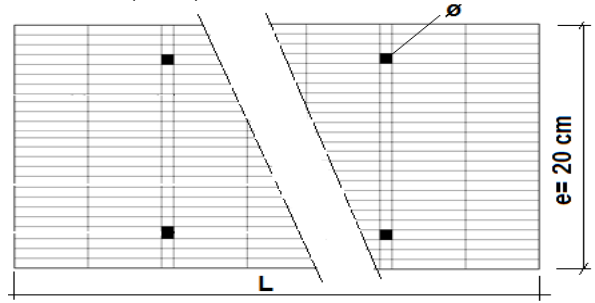

Figure 1. Discretization of walls

The two other walls, "Mu (12) 20" and "MuCH (12) 20" have a section of $20 \times 350$ cm². They are armed with steel of diameter Ø 12 spaced with 20 cm. Mu (12) 20 is submitted to its ultimate loading (moment and load). The Wall "MuCH (12) 20" has the same mechanical loading (ultimate moment and ultimate compressive load) that Mu 20. For the thermal loading, "Mu (12) 20" and "MuCH (12) 20" are exposed to a fire ISO834 [3] on one side. The floor height (H) is equal to 2.86 m for the four walls.

TABLE II. GEOMETRICAL CHARACTERISTICS AND LOADING OF CONSIDERED WALLS

| Walls | H(m) | L (cm) | e (cm) | Ø (mm) | thermal loading | mechanical loading |
|---|---|---|---|---|---|---|
| Mu 20 | 2,86 | 470 | 20 | 10 | ISO834 on one side | N-ult,M-ult of (Mu20) |
| MuF 20 | 2,86 | 470 | 20 | 10 | ISO834 on two sides | N-ult,M-ult of (MuF20) |
| Mu (12)20 | 2,86 | 350 | 20 | 12 | ISO834 on one side | N-ult,M-ult of (Mu (12)20) |
| MuCH (12)20 | 2,86 | 350 | 20 | 12 | ISO834 on one side | N-ult,M-ult of (Mu 20) |

## III. THERMAL ANALYSIS

### A. Basic equation

In the software Safir, the heat flux exchanged between a boundary and the hot gas in a fire compartment can be modeled according to the recommendation of Eurocode 1 with a linear convective term and radiation term, see Equation 5.

$$q_n = h(T_g - T_s) + \sigma \varepsilon^* (T_g^4 - T_s^4) \quad (6)$$

$\sigma$ : Stefan-Boltzman coefficient, $5.67 \times 10^{-8}$
$\varepsilon^*$ : relative emissivity of the material
h : coefficient of convection, $w/m^2$-K
$T_g$ : temperature of the gas, given in the data as a function of time, K
$T_s$: temperature on the boundary of the structure, K

### B. Temperatures in the wall « MuF20 »

In this numerical study, the thermal analysis is a prerequisite for any result, so we start firstly by determining the temperatures at each point of the walls by using code "SAFIR". This code is based on norms [7] and [8]. We cite two cases of walls exposed to fire (MuF 20 and Mu (12) 20).

In the case of the wall "MuF 20" which is exposed to fire(in red) in two faces "Fig. 2", at failed time t = 8940sec or 149min (2.43 h), we get a temperature between 900.78 and 1048.50 ° C at the surfaces in contact with the fire. Away from both faces exposed to fire, temperature decreases to 457.60 °C.

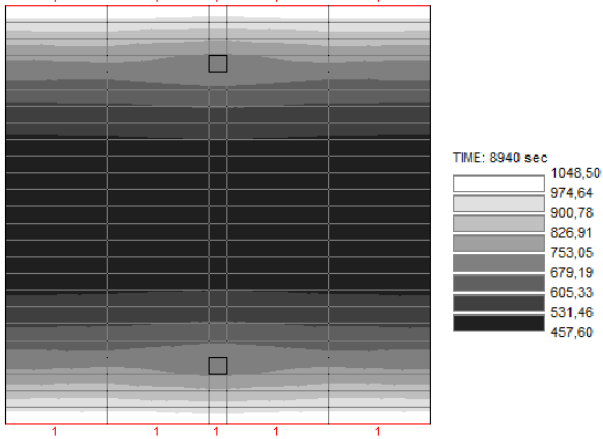

Figure 2. Temperatures of wall « MuF 20 » at failed time Units

C. *Temperatures in the wall « Mu (12)20 »*

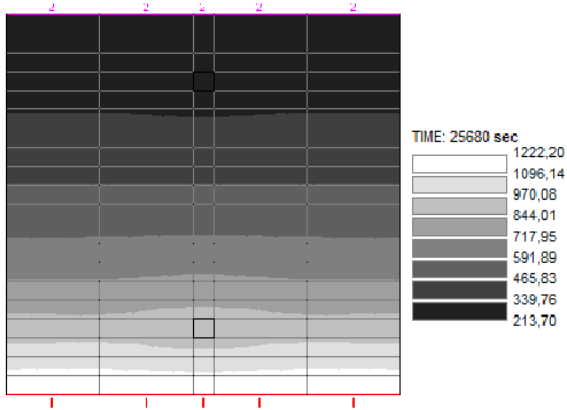

Figure 3. Temperatures of wall « Mu (12)20 » at failed time

The results obtained from the numerical study concerning variations in temperature at the ruin time in the concrete section are in "Fig. 3". For the failed time (25680s) or 7h 13min, observed temperature of the face exposed to fire (number 1) varies between 970,08 and 1222.20 °C. In the side who is not exposed to fire (number 2), the temperature is 213,70 °C at the failed time. Of course after a long period, the temperature rises considerably.

IV. MECHANICAL ANALYSIS

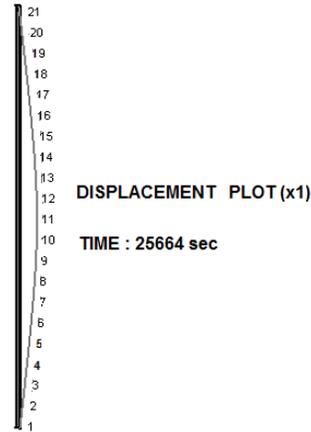

Figure 4. Appearance of Mu (12) 20 at the failed time

The mechanical data file is dependent on thermal analysis for the use of the elements temperatures in function
of time. The file in question contains the dimensions of the wall (height and width), the number of nodes is equal to 21. The number of the beam element according to the discretization is taken equal to 10, each element contains 3 nodes. Mechanical loading is represented by a normal force and moment for each wall, the calculation is performed for a
strip of 20 cm. Figure 4 shows the appearance of the wall *Mu (12) 20* at failed time (t = 25664 sec).

We note from Table III, that Mu (12) 20 has a better fire behavior compared to other walls, because of its good rigidity. MuF20 is identical to Mu 20; however MuF20 is exposed to fire at two sides which explains the good fire behavior of Mu 20 compared to the behavior of MuF 20.

TABLE III. FIRE RESISTANCE OF CONSIDERED WALLS

| Wall | coating (cm) | Height (m) | Ø (mm) | span (m) | Depth (m) | M-ult (t.m) | N-ult (t) | Rf (min) |
|---|---|---|---|---|---|---|---|---|
| Mu 20 | 2,4 | 2,86 | Ø 10 | 4,7 | 0,2 | 2,26 | 41,2 | 179,78 |
| MuCH (12)20 | 3 | 2,86 | Ø 12 | 3,5 | 0,2 | 2,26 | 41,2 | 278,93 |
| Mu (12)20 | 3 | 2,86 | Ø 12 | 3,5 | 0,2 | 0,14 | 26,32 | 427,75 |
| MuF 20 | 2,4 | 2,86 | Ø 10 | 4,7 | 0,2 | 2,26 | 41,2 | 148,76 |

A. *Displacement and strain of Mu (12)20*

In Figure 5, the curve represents the horizontal displacement of the wall "Mu (12) 20." There is a positive evolution (dilatation) during the exposition to fire. The maximum displacement of the node 11 (middle of the wall) at the collapse is 10 cm after a period of t = 25500sec (7h), which is representing 50% of the wall thickness. This displacement is the largest (buckling phenomenon). Given that Mu (12) 20 with section (20x 350) is exposed to fire on one side and mechanically loaded with an ultimate load of 15,040 N and an ultimate moment of 80 N.m according to [2] [9]. We can say that this wall has a good fire resistance.

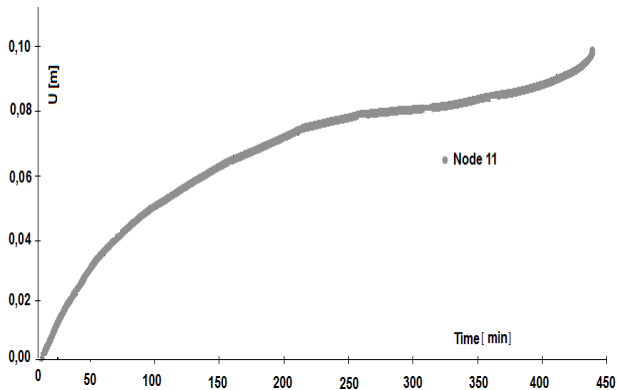

Figure5. Horizontal displacement of Mu (12)20 at half height

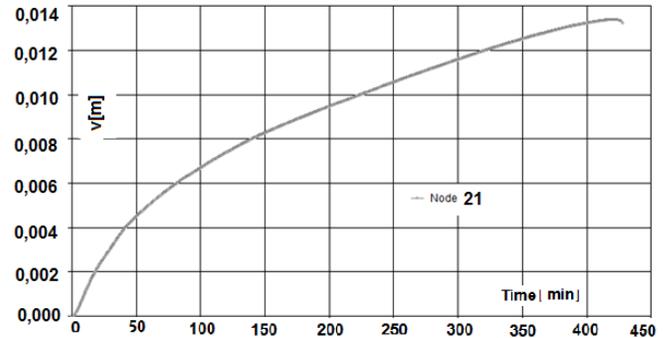

Figure 6. Vertical displacement of Mu (12)20 at the upper end

In the case of the node 21 "Fig. 6" located at the upper extremity, the node 21 presents the maximum vertical displacement in sight of the boundary conditions. The vertical displacement is positive and equal to 1, 4 cm (there is an expansion due to thermal loading). This displacement is followed by the collapse of the wall at 25500sec (7h).

B. *Displacement and strain of Mu20*

The mechanical analysis shows that Mu20 deforms with increasing temperature and with time. Curve of node 11"Fig. 7" represents a positive evolution throughout the time of exposition to fire.

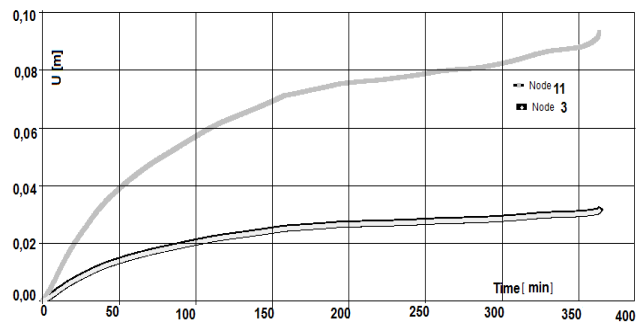

Figure 7. Horizontal displacements of nodes 3 and 11

The maximum displacement at the collapse is 9cm, given that Mu20 (20x470) is exposed to fire on one side and mechanically loaded with a force (17532 N) and a moment (961,7 N.m). But the node 3 has a small displacement, equal to 3cm.

C. *Displacement and strain of MuF 20*

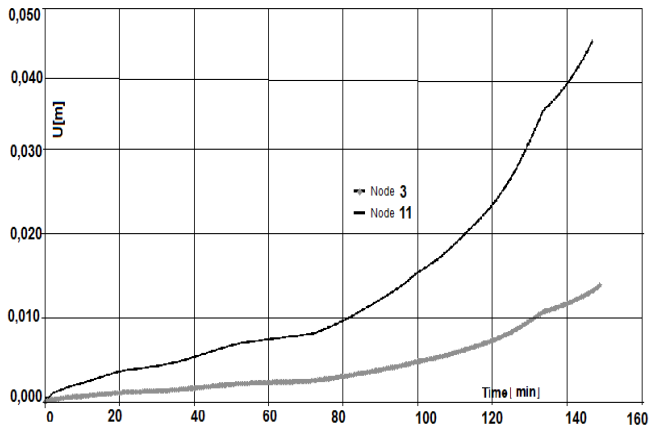

Figure 8. Horizontal displacement of the nodes 3 and 11

"Fig.8" shows the horizontal displacements of nodes 3 and 11. MuF20 (20x470) is exposed to fire on both sides, in the case of node 11, whose curve has a greater displacement reaching 4,5cm in an estimated time of 8925sec (149 min). but node 3 has a small displacement equal to 1, 5 cm.

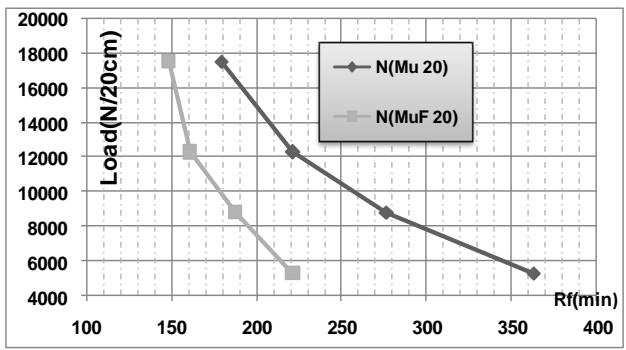

Figure 10. Fire resistance of walls « Mu 20 » and « MuF 20 » depending on the load

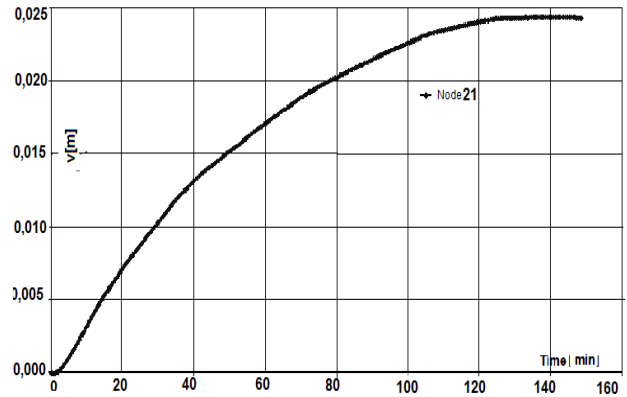

Figure 9. Vertical displacement of wall MuF20 at the upper end

The Wall "MuF20" at its upper end (node 21), underwent dilatation (vertical displacement) of 2,5 cm after an estimated ruin time of 8900sec (148 min) "Fig. 9". We note that this displacement is important compared to vertical displacements of previous walls, because MuF 20 is exposed to fire according two sides, its dilatation is considerable. In addition, loading has a considerable effect on the walls in case of fire. The fire acts indirectly on the structures (reinforced concrete walls), it destroys the mechanical properties of materials (concrete, steel), so that they become incapable of supporting the loads.

The curves obtained in "Fig.11"; show the fire resistances of two walls exposed to fire on one side, and submitted identically to different rates of mechanical loading. These two walls does not have the same dimensions, but are subject to the same mechanical loading and to the same thermal loading (ISO834), as it was mentioned previously.
Their sections are respectively, for Mu 20: $20\times470cm^2$ and for MuCH (12) 20: $20\times350$ cm$^2$. We note that the fire resistances of these two types of walls are considerably higher than preceding walls (Figure 10). We observe that MuCH (12) acts better than Mu 20. The section of MuCH (12)20 is lower than to that of Mu 20, thus his stress resistance ($\sigma$=N/S) is greater than the stress resistance of Mu 20. Otherwise the section of reinforcement (Ø12) of MuCH (12) 20 is greater than the section of reinforcement (Ø10) of Mu 20.

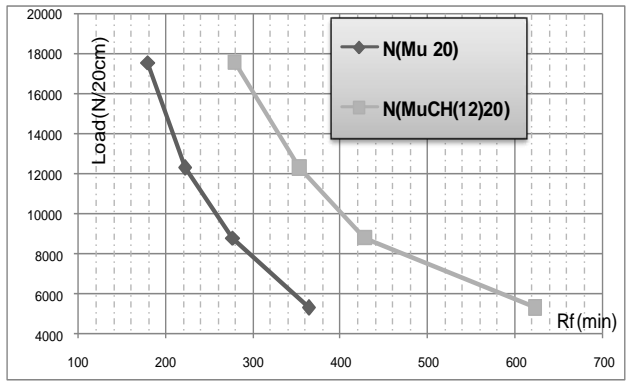

Figure 11. Fire resistance of walls « Mu 20» and «MuCH (12)20 » depending on the load

VI. CONCLSIONS

- Mu 20 has a better fire behavior than MuF20 (exposed to fire at two sides) despite of their similarities "Fig.12".
- The walls "Mu (12) 20 and" MuCH (12) 20 " are similar but the mechanical loading of the first is smaller than the second which gives that the resistance of Mu (12) 20 is equal to the double of the resistance of MuCH (12) 20.
- We note that the sizing has a significant effect on fire resistance, as well, Mu 20 with a section of 20x470 cm$^2$ is less resistant than Mu (12) 20 having a section : 20x350 cm$^2$. It is recommended to forecast column, when length of the firewall exceed 3.5 m, to reduce the span wall.
- The fire resistances of walls considered are close to the fire resistances given by the norms [table1] [5] [6]. On the other hand the displacements of walls are in accordance with the appearance of the curves founded by (A Nadjai, 2006).
- We note that mechanical loading has a considerable effect on the walls in case of fire, experimental results of numerous researches of structures studied (for example, in university of Liege in BELGIUM), have previously demonstrated that the fire acts indirectly on the structures (in our case the reinforced concrete walls). The fire destroys the mechanical properties of materials (concrete, steel), so that they become unable to bearer the mechanical load.
- In order to know the impact of dimensioning, more precisely of the span wall in case of fire we considered two walls (Mu 20 and MuCH (12) 20) not having the same dimensions exposed every two, to the same mechanical loading and to the same thermal loading. We

V. COMPARISON OF CONSIDERED WALLS

In "Fig.10" the curves show the fire resistance of two walls exposed at fire; the first on two sides (grey curve) and the second (black curve) on one side, considering four rates of loading (100%, 70%, 50% and 30%).

We find that the resistance of Mu 20 who was exposed on one side is larger than the resistance of MuF 20 which was exposed to fire on both sides. We also find that mechanical

The Analysis of these results shows that the increasing of the span wall causes a reduction in the fire resistance.
In addition, this analysis justifies well the use of reinforced concrete walls, to limit the spread of fire from one structure to another structure nearby, since the resistances obtained are considerable (10 hours). Finally, we deduce that MuCH(12) 20 has a better fire behavior (stop fire) because it has a good fire resistance which reaches 10 hours.

observed that the fire resistance of the wall with the little span is considerably higher than that fire resistance wall with the great span. We conclude that a significant span, more than 3 m is unfavorable for the firewall, since it leads to a reduction in fire resistance.
- Walls studied have appreciable fire resistances, which justifies well the use of reinforced concrete walls (firewall), to limit the spread of fire from one structure to another structure nearby. Particularly "Mu (12) 20» has an appropriate size, allowing it to play the role of a firewall, because it has better fire resistance and good rigidity.
- Furthermore, it would be interesting to carry out an experimental study on the walls considered to complete this work.

ACKNOWLEDGMENT

The work presented was possible with the assistance of Professor Jean Marc Franssen and Mr. Thomas Gernay which we thank gratefully.

TABLE IV. NOMENCLATURE

| | |
|---|---|
| Mu 20: | Reinforced concrete wall with a thickness equal to 20cm and a span equal to 470cm (reinforcement :Ø10), this wall was exposed to fire on one side. |
| MuF 20: | Reinforced concrete wall with a thickness equal to 20cm and a span equal to 470cm (reinforcement :Ø10), this wall was exposed to fire on two sides. |
| Mu (12) 20: | Reinforced concrete wall with a thickness equal to 20cm and a span equal to 350cm (reinforcement :Ø12), this wall was exposed to fire on one side. |
| MuCH (12) 20: | Reinforced concrete wall with a thickness equal to 20cm and a span equal to 350cm (reinforcement: Ø12), this wall was exposed to fire on one side. In this case, we use the ultimate mechanical loading of wall Mu 20. |
| Firewall | Reinforced concrete wall intended to limit the spread of fire from a structure to another nearby. |
| N-ult: | Ultimate compressive load. |
| M-ult: | Ultimate moment. |
| $\sigma$ : | Stress. |
| Rf: | Fire resistance. |
| H: | Height. |
| h: | Hour. |
| L: | Span of wall. |
| e: | Thickness of wall. |
| N: | Compressive load. |
| S: | Area of section. |
| Ø: | Diameter of used steel. |